\documentclass[epj]{svjour}
\usepackage{graphics}
\usepackage{lipsum}
\usepackage{mathtools}
\usepackage{amsmath}
\usepackage{cuted}
\begin{document}
\title{\textbf{New static cylindrical symmetric solutions in GR for unfamiliar EoS parameter values via Noether symmetry}}
\author{Işıl Başaran Öz\inst{1}\thanks{\emph{e-mail:} ibasaranoz@gmail.com (corresponding author)}\and Kazuharu Bamba\inst{2}\thanks{\emph{e-mail:}  bamba@sss.fukushima-u.ac.jp}}
\institute{Department of Physics, Faculty of Science, Istanbul University 34134, Istanbul Turkey  \and Division of Human Support System, Faculty of Symbiotic Systems Science, Fukushima University, Fukushima 960-1296, Japan}
\date{Received: date / Revised version: date}
\abstract{
The solutions for the field equations of $f(R)$ gravity are investigated in static cylindrically symmetric space-time. Conserved quantities of the system, as well as unknown functions, can be determined with the help of the Noether symmetry method. In this article, some unknown values of the equations of state parameter (EoS) have emerged as a result of the constraints obtained by analyzing the Noether symmetry equations for the $f(R)=f_0 R$ case. Consequently, several new exact solutions have been found for cases of General Relativity in static cylindrically symmetrical space-time for the non-dust matter.
} 
\maketitle
\section{Introduction}
\label{intro}

One of the most important discoveries in cosmology is the finding that our universe is accelerating the expansion as a result of the supernova Type Ia observations made by two independent groups at the end of the last century \cite{Perlmutter1998,Riess1998}. Although Einstein's theory of General Relativity (GR) is taken as the basis to explain the structure of the universe, it falls short of explaining this accelerating expansion \cite{Weinberg1972}. However, the cosmological constant that Einstein put in his equations to make it static and which he later thought was his biggest mistake is today considered vacuum energy for dark energy. The $\Lambda$CDM model created by taking vacuum energy together with cold dark matter (CDM) is accepted as the concordance model of cosmology \cite{padmanabhan2007}. Although this model is quite compatible with observations, it cannot solve the problems such as coincidence, fine-tuning, etc. \cite{Copeland2006,Durrer2008}. Therefore, in order to explain these phenomena, modified gravitation theories have been starts studied by scientists \cite{Clifton2012,Bamba2012,Capozziello2011}. These modifications are based on changing the geometry and/or matter parts of the GR. One of the most basic modifications of the geometry part is the $f(R)$ gravity theory, which is based on taking the $f(R)$ function in GR instead of the $R$ Ricci scalar \cite{Copeland2006,Nojiri2010,Nojiri2017,Sotiriou2008,DeFelice2010}. For the application to the cosmic acceleration in the early universe, e.g., Ref. \cite{Starobinsky1980,Carroll2003,Nojiri2003,Capozziello2003}. The thermodynamics of the apparent horizon in $f(R)$ gravity has been studied in \cite{Bamba2009,Bamba2010a}. Curvature singularity that occurs in the process of stellar collapse in $f(R)$ gravity has been investigated in \cite{Bamba2011}. Future transitions of the phantom divide line $w_\mathrm{de}=-1$ of the equation of state (EoS) parameter for dark energy have shown that the sign of the time derivative of the Hubble parameter will change from this negative to positive in the current applicable gravity models $f(R)$ in \cite{Bamba2010b}. Reconstruction of $f(R)$ gravity models with bounce cosmology is performed in \cite{Bamba2013}.

Cylindrical symmetrical space-time plays an important role in studying the homogeneous but anisotropic structure of the universe on a large scale. Besides, cylindrical symmetric spacetimes are widely studied within the framework of both GR and modified theories such as $f(R)$, $f(T)$, etc. \cite{Trendafilova2011,Sharif2012b,Houndjo2012,FarasatShamir2014}. Also, this background is suitable for the study of compact objects \cite{Sharif2014,Bhatti2017,Yousaf2020}. Additionally, it is frequently used for string theory and wormhole studies, as it has an important geometric structure \cite{Momeni2017,Azadi2008,Bronnikov2019}. Therefore, conducting cosmological studies in the context of cylindrical symmetrical space-times would be valuable because of its potential to bring different parts of the puzzle together.

There are many studies in the literature for its applications to cosmic acceleration in the early universe, e.g., Ref. \cite{Starobinsky1980,Carroll2003,Nojiri2003,Capozziello2003}. The thermodynamics of the apparent horizon in $f(R)$ gravity has been studied \cite{Bamba2009,Bamba2010a}. Curvature singularity that occurs in the process of stellar collapse in $f(R)$ gravity has been investigated \cite{Bamba2011}. Future transitions of the phantom divide line $w_\mathrm{de}=-1$ of the equation of state (EoS) parameter for the $f(R)$ gravity have shown that the sign of the time derivative of the Hubble parameter will change from this negative to positive see Ref. \cite{Bamba2010b}. Reconstruction of $f(R)$ gravity models with bounce cosmology is performed \cite{Bamba2013}.

In the literature, there are $f(R)$ and some modified gravitational theories solutions studied by the Noether symmetry approach for spherical symmetric space-time \cite{Capozziello2007,Capozziello2012,Bahamonde2018}. Some black hole solutions investigated under $f(R)$ theory using Noether symmetry for BTZ spacetime are also available in the literature \cite{Camci2020,Darabi2013}. Unlike all these studies, in this work, we wanted to examine how would be at the solutions a cosmological scale and with a cylindrical symmetrical background, rather than a compact object.

The article is organized as follows. In Sec. 2, we construct the theory of $f(R)$ gravity with a static cylindrical symmetric background, and then in Sec. 3, we explain the Noether symmetry approach. In Sec. 4, we present the Noether symmetries to these theories of gravity and exact solutions through their first integrals. Finally, in Sec. 5, we summarize the findings through this study.
\section{Field Equations}
\label{sec:FEq}

In this section, we discuss the metric $f(R)$ theory. The $f(R)$ gravitational theory is an interesting and relatively simple alternative that can be considered instead of the GR theory. We determine the matter Lagrangian form for the static cylindrical symmetric space-time and adding to the action write Lagrangian of the theory, and then present the field equations. For an introduction to Metric $f(R)$ gravity see \cite{Sotiriou2008,Azadi2008,Sawicki2007,Sharif2012a}, comprehensive analysis of all versions of $f(R)$ gravity and for alternative gravity theories  also see \cite{Clifton2012,Nojiri2010,Nojiri2017,DeFelice2010}. 
Beginning from the action $GR$ and adding the $f(R)$ term rather than the R term and add also a matter term $S_m$, the $4-$dimensional total action for $f(R)$ theory of gravity takes the form:

\begin{equation}\label{actf(R)}
\mathcal{S}=\int{d^4x \sqrt{-g}\left[\frac{1}{2 \kappa}f(R)+\mathcal{L}_m\right]},
\end{equation}

where $\kappa=8\pi G$ and $\mathcal{L}_m$ is Lagrangian related to matter content of any kind in the universe. $f(R)$ is a function dependent on the scalar curvature $R$, while $ f(R)=R $ the theory is clearly reduced to the General Relativity \cite{Sotiriou2008}.
The static cylindrical symmetric space-time metric is as follows \cite{Momeni2009,Shamir2014}.

\begin{equation}\label{sss-metric}
  ds^2=A(r)dt^2-dr^2-B(r)( d\theta^2+\alpha^2 dz^2),
\end{equation}

where $ A(r) $ and $ B(r) $ are the metric coefficients of taken depending on the radial coordinate  $ r $.
Using the Lagrange Multiplier Method, which is generally adopted for higher-order theories, then applying the method of integration by parts, the point-like Lagrangian is derived as follows \cite{Capozziello1994}.

\begin{strip}
\begin{equation}\label{lag}
\mathcal{ L}=-\frac{\alpha f_{R}}{\sqrt{A}}A'B'-\alpha f_{RR}\frac{B}{\sqrt{A}}A'R'-\frac{\alpha f_{R}}{2B}\sqrt{A}B'^2-2\alpha f_{RR}\sqrt{A}B'R'+\alpha \sqrt{A}B[f-R f_{R} -\kappa \rho_{0} A^{-\frac{1+w}{2w}}], \qquad (w \neq 0)
\end{equation}
\end{strip}

where $\mathcal{L}_m=-\rho$, ($'$)denotes the derivative with respect to $r$ and $f_R\equiv df/dR$, $f_{RR}\equiv d^2f/dR^2$. Also, configuration space of the (\ref{lag}) Lagrangian is $Q=\{A, B, R\}$.

For the metric (\ref{sss-metric}), the matter density is obtained as $\rho= \rho_{0} A^{-\frac{1+w}{2w}}$. Here, $w$ is the equation of state parameter, which $w=\frac{1}{3}$ for the radiation dominant case and $w=0$ for the mater dominant case. When $w=-1$, dark energy is dominant which refers to the negative pressure fluid. Besides, it is known that for a gas composed of cosmic strings $w=-\frac{1}{3}$ and for the stiff matter $w=1$. Here we examine the non-dust solutions in line with the constraint evident from the Lagrangian.
From the variation with respect to the configuration space elements $A$, $B$ and $R$ for the Lagrangian (\ref{lag}), the field equations of the theory of gravity obtained are as follows:

\begin{strip}
\begin{eqnarray}\label{alan1}
\frac{f_{R}}{4} \, \left(2 \frac{A''}{A} +2\frac{A'B'}{AB}-\frac{A'^2}{A^2}\right)-\frac{f}{2}-\left(f_{RR} R'\frac{B'}{B}+f_{RR}R''+f_{RRR}R'^2\right)+\frac{1}{w}\kappa \rho_0A^{-\frac{1+w}{2w}}=0,
\end{eqnarray}
\begin{eqnarray}\label{alan2}
-\frac{f_{R}}{4} \, \left(2 \frac{A''}{A}-\frac{A'^2}{A^2}+ 4\frac{B''}{B}-2\frac{B'^2}{B^2}\right)+\frac{f}{2}+\frac{f_{RR} R'}{2}\left(\frac{A'}{A}+2\frac{B'}{B}\right)-\kappa  \rho_0A^{-\frac{1+w}{2w}}=0,
\end{eqnarray}
\begin{eqnarray}\label{alan3}
-\frac{f_{R}}{4} \, \left(2 \frac{B''}{B}+\frac{A'B'}{AB}\right)+\frac{f}{2}+\frac{f_{RR} R'}{2}\left(\frac{A'}{A}+\frac{B'}{B}\right)+f_{RR}R''+f_{RRR}R'^2-\kappa \rho_0A^{-\frac{1+w}{2w}}=0.
\end{eqnarray}
\end{strip}

These equations are fourth-order nonlinear differential equations and it is not easy to find an exact solution without making any approach. In this study, we use the Noether symmetry approach to find a solution.
\section{Noether Gauge Symmetry}
\label{sec:NGS}

Noether symmetry approach has been widely used in the literature to find exact solutions, particularly of modified gravity theories \cite{Capozziello2007,Capozziello1996,Capozziello2008,Hussain2011,Jamil2011,Kucuakca2011,Vakili2008,Shamir2017,Oz2017} . It opens the way of the solution by decreasing the degrees of freedom of the dynamic system and/or determining the unknown functions of the system. In a sense, the existence of a Noether symmetry is a kind of choice rule. The outline of the Noether symmetry method is given below.

The Lagrangian-related Noether vector $X$ and the first prolongation vector field $X^{[1]}$ can be built as follows:

\begin{equation}\label{NGSvec}
X \mathcal{L}=\xi(\tau,q^k)\frac{\partial\mathcal{L}}{\partial \tau}+\eta^i(\tau,q^k)\frac{\partial\mathcal{L}}{\partial q^i}, \nonumber
\end{equation}
\begin{equation}\label{NGS1vec}
X^{[1]} \mathcal{L}=X \mathcal{L}+\dot{\eta}^k(\tau, q^l, \dot{q}^l)\frac{\partial\mathcal{L}}{\partial \dot{q}^k }. \nonumber
\end{equation}

Here it is defined as $ \dot{\eta}^k(\tau, q^l, \dot{q}^l)=D_\tau\eta^k -\dot{q}^kD_\tau\xi$, and $ D_\tau=\partial/\partial \tau +\dot{q}^k\partial/\partial q^k $ is the total derivative operator.

If there is a $ G (\tau, q ^ k) $ function for any $ \mathcal {L} $ Lagrange function and the following Noether symmetry condition is satisfied,

\begin{equation}\label{NSkoş}
X^{[1]} \mathcal{L}+\mathcal{L}(D_\tau\xi)=D_\tau G.
\end{equation}

There is a conserved quantity that belongs to the system of the equation that is expressed as follow

\begin{equation}\label{FirstInt}
I=-\xi (\dot{q}^i \frac{\partial \mathcal{L}}{\partial \dot{q}^i} -\mathcal{L})+\eta^i\frac{\partial L}{\partial \dot{q}^i}-G.
\end{equation}

This expression is important in that it provides a solution for the system of differential equations of the theory.
When the generalized coordinates $ q^i $ are taken as the dynamic variables of the extended gravitational theory considered, the conservative quantities associated with the theory of gravity can be found using the Noether symmetry approach described above. Thus, it can be possible to obtain a new exact solution for the corresponding gravitational theory model.
\section{Static cylindrically symmetric solutions}
\label{sec:SOL}

Noether symmetry condition (\ref{NSkoş}) gives the following system of differential equations:

\begin{strip}
\begin{eqnarray}\label{ngs}
& & \xi_{,A} =0, \qquad \xi_{,B} =0, \qquad \xi_{,R} =0, \qquad  \alpha \,(  f_{R}
\eta^2_{,r} + B f_{RR} \eta^3_{,r})+\sqrt {A}G_{,A}=0,
\nonumber \\
& &  \alpha\,\left[ f_{R}\left(\eta^1_{,r}+{\frac {A\eta^2_{,r}}{B}} \right) +2 \, A f_{RR} \eta^3_{,r} \right]+\sqrt {A} G_{,B} =0,\qquad  \alpha \, f_{RR} \, ( B\eta^1_{,r} +2\,A\eta^2_{,r} ) +\sqrt {A}G_{,R} =0,
\nonumber \\
& &f_{R} \eta^2_{,A} +B f _{RR} \eta^3_{,A} =0,   \qquad  \frac { f_{R} \eta^1  }{2 A}-\frac { f_{R} \eta^2 }{B} +  f _{RR}\eta^3 +2\,\frac {B f _{R}\eta^1_{B}}{A}+2\, f _{R} \eta^2_{B} +4\,f _{RR} B\eta^3_{B}- f _{R} \xi_{r}=0,
\nonumber \\
& &  -\frac{f_{R} \eta^1 }{2A}+ f _{RR} \eta^3 + f _{R} \eta^1_{,A} + f _{R}\frac {A }
{B}\eta^2_{,A}+2\,A  f _{RR}\eta^3 _{,A} + f _{R} \eta^2_{,B} + f _{RR} B \eta^3 _{,B} - f _{R}\xi _{,r} =0,
\nonumber \\
& & -\frac{f_{RR} \eta^1 }{2A}+\frac{f_{RR} \eta^2 }{B}+ f_{RRR} \eta^3 +\frac{f_{R}\eta^2 _{,R} }{B}+ f _{RR}\eta^3 _{,R} + f _{RR} \eta^1_{,A}+2\,\frac
{A f _{RR} \eta^2 _{,A}}{B}- f _{RR} \xi_{,r} =0,
\nonumber \\
& &  \frac {  f_{RR} \eta^1 }{A}+2\, f _{RRR} \eta^3+\frac{f_{R} \eta^1}{A}+\frac{f _{R} \eta^2_{,R} }{B}+2\, f _{RR} \eta^3_{,R} +\frac {B f _{R} \eta^1 _{,B} }{A}+2\, f _{RR} \eta^2 _{,B} -2\, f _{RR} \xi _{,r} =0,
\nonumber \\
& & \left( -f
	+Rf_{R} +\kappa\,\rho_{{0}} {A}^{-\frac {1+w}{2w}} \right)\left[\frac {\alpha\,B}{2\sqrt {A}}\eta^1+\alpha\,\sqrt {A}\eta^2+ \alpha\,\sqrt {A}B\xi_{,r}\right]  - \frac {	\alpha\,B\kappa\,\rho_{{0}} (1+w) }{\sqrt {A}w}{A}^{-\frac {1+w}{2w}} \eta^1  \nonumber \\
& &   +\alpha\,\sqrt {A}BR f_{RR} \eta^3  +G_{,r} =0, \qquad  B\eta^1 _{,R} +2\,A\eta^2 _{,R} =0.
\end{eqnarray}
\end{strip}

Solutions have been studied under various conditions for this system of equations. Generally, a different symmetry is not observed for the case where $f(R) = R$, ie for the GR lagrangian. One of the important points of this study is that when examining equations (\ref{ngs}), the existence of excess symmetry was observed for the values of the equation of state parameter, $ w=-1/4$ and $ w=1/5 $. These situations are discussed separately below.

{{\bf  (i)}:} Noether symmetry vector components under conditions $ f (R) = f_0 R $, $ w = -1/4 $, $ f_0, \rho_0> 0 $ are as follows:

\begin{eqnarray}\label{ç_ii}
& &   \xi=c_1r+c_2, \nonumber \\
& &  \eta^1=-c_1\frac{4A}{3}-\frac{A}{2B^{3/4}}(c_3r+c_4),\nonumber \\
& & \eta^2=c_1\frac{5B}{3}+B^{1/4}(c_3 r+c_4),\nonumber \\
& &  G=-c_3 2\alpha f_0 \sqrt{A}B^{1/4} +c_5.
\end{eqnarray}

Thus, Noether symmetries are obtained as follows

\begin{eqnarray}\label{s_ii}
& &   \textbf{X}_1=\partial_r, \nonumber \\
& & \textbf{X}_2=r\partial_r-\frac{4A}{3}\partial_A+\frac{5B}{3}\partial_B, \nonumber \\
& & \textbf{X}_4=-\frac{A}{2B^{3/4}}\partial_A+B^{1/4}\partial_B,\nonumber \\
& & \textbf{X}_3=-\frac{A}{2B^{3/4}}r\partial_A+B^{1/4}r\partial_B, \nonumber \\
& & G=- 2\alpha f_0 \sqrt{A}B^{1/4}.
\end{eqnarray}

For each Noether symmetry, there is a Noether integral, ie a conservative quantity. These are determined as follows

\begin{eqnarray}\label{int_ii}
& & I_1=-E_{\mathcal{L}}=0, \nonumber \\
& & I_2=-\frac{1}{3}\alpha f_0\sqrt{A}B\left(\frac{B'}{B}+5\frac{A'}{A}\right),\nonumber \\
& & I_3= -\frac{1}{2}\alpha f_0\sqrt{A}B^{1/4}\left[4-r\left(\frac{B'}{B}+2\frac{A'}{A}\right)\right], \nonumber \\
& & I_4=\frac{1}{2}\alpha f_0\sqrt{A}B^{1/4}\left(\frac{B'}{B}+2\frac{A'}{A}\right).
\end{eqnarray}

When the above Noether integrals are examined, it is clearly seen that there are relations between them as follows

\begin{eqnarray}\label{}
& &  I_3-rI_4=2\alpha f_0\sqrt{A}B^{1/4},\nonumber \\
& &  I_4=\frac{rI_4-I_3}{4}\left(\frac{B'}{B}+2\frac{A'}{A}\right),\nonumber \\
& &  BA^2=k_0(rI_4-I_3)^4.
\end{eqnarray}

From these relations, the metric coefficients $ A (r) $ and $ B (r) $ can be easily obtained as follows.

\begin{eqnarray}\label{bulg_ii}
& &A(r)=\left[-\frac{3I_2}{10k_0(rI_4-I_3)^3\alpha f_0I_4}+k_1(rI_4-I_3)^2\right]^{-2/3},\nonumber \\
& &B(r)=k_0(rI_4-I_3)^4\frac{1}{A(r)^2},
\end{eqnarray}

where $\rho_0 =\frac{40}{3\kappa}f_0 k_1 I_4^2 $.
\begin{figure}
\resizebox{0.40\textwidth}{!}{\includegraphics{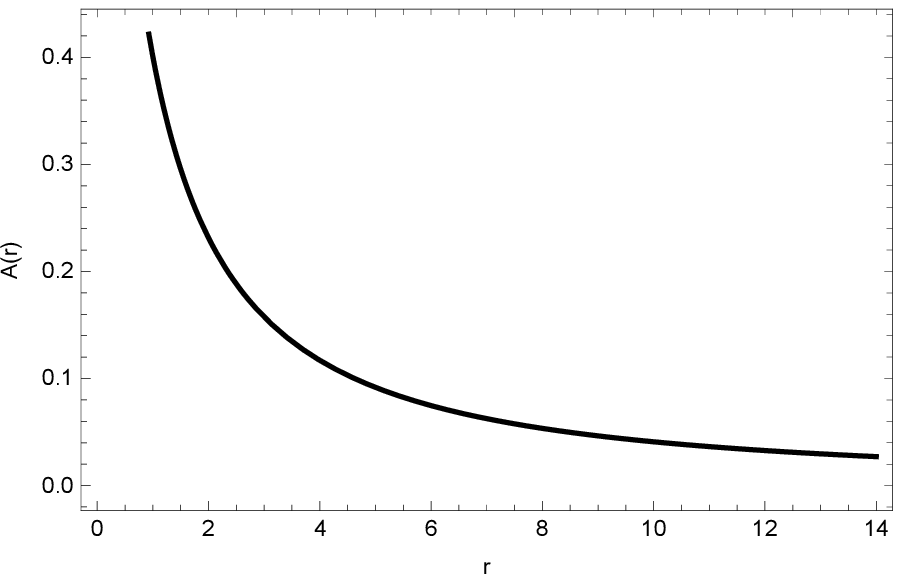}}
\caption{{Metric coefficient $A(r)$ as a function of the radial coordinate $r$ for the case of $w=-1/4$ with $\alpha=2$, $f_0=1/3$, $a_0=1$, $k_0=k_1=1$, $I_3=-1$ and $I_2=I_4=1$.}}
\label{fig:A(r)}       
\end{figure}

\begin{figure}
\resizebox{0.40\textwidth}{!}{%
  \includegraphics{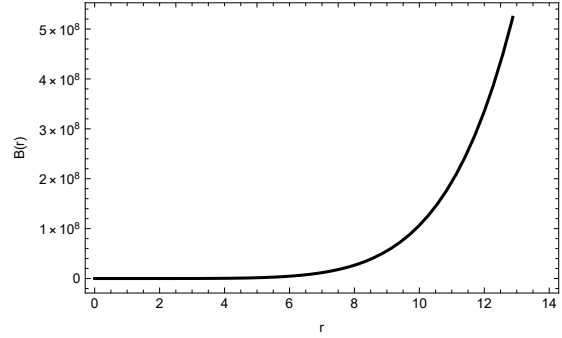}}
\caption{{Metric coefficient $B(r)$ as a function of the radial coordinate $r$ for the case of $w=-1/4$ with $\alpha=2$, $f_0=1/3$, $a_0=1$, $k_0=k_1=1$, $I_3=-1$ and $I_2=I_4=1$.}}
\label{fig:B(r)}       
\end{figure}

Symmetries under these constraints allow us to find the exact solution of the equations of the theory. The critical point here is that the EoS parameter points to a specific value of $ w = -1/4 $, which also corresponds to the Quintessence value range.

{{\bf  (ii)}:} Under conditions $f(R)=f_0 R $, $w=1/5$, $f_0, \rho_0 >0 $ Noether symmetry vector components under conditions $  f(R)=f_0 R $, $  w=1/5 $, $ f_0, \rho_0 >0 $ are as follows
\begin{eqnarray}\label{ç_iii}
& &   \xi=c_1\frac{r^2}{2}+c_2r+c_3, \nonumber \\
& & \eta^1=c_1\frac{2A}{3}r+c_2\frac{2A}{3},\nonumber \\
& & \eta^2=c_1\frac{2B}{3}r+c_2\frac{2B}{3}, \nonumber \\
& & G=-c_1\frac{4}{3} \alpha f_0 \sqrt{A}B +c_4.
\end{eqnarray}

Thus, Noether symmetries are obtained as follows

\begin{eqnarray}\label{s_iii}
& &    \textbf{X}_3=\partial_r, \nonumber \\
& & \textbf{X}_2=\frac{r^2}{2}\partial_r+\frac{2A}{3}r\partial_A+\frac{2B}{3}r\partial_B, \nonumber \\
& & \textbf{X}_3=r\partial_r+\frac{2A}{3}\partial_A+\frac{2B}{3}\partial_B,\nonumber \\
& & G=-\frac{4}{3} \alpha f_0 \sqrt{A}B.
\end{eqnarray}

A Noether integral for each Noether symmetry is computed as follows

\begin{eqnarray}\label{int_iii}
& &    I_1= -E_{\mathcal{L}}=0,\nonumber \\
& & I_2=\frac{2}{3}\alpha f_0\sqrt{A}B\left[2-r\left(2\frac{B'}{B}+\frac{A'}{A}\right)\right], \nonumber \\
& & I_3=-\frac{2}{3}\alpha f_0\sqrt{A}B\left(2\frac{B'}{B}+\frac{A'}{A}\right).
\end{eqnarray}

With the Noether integrals above, the following relation is obtained

\begin{equation}\label{bulg_iii}
 \sqrt{A}B= \frac{3}{4\alpha f_0}(I_2-rI_3).
\end{equation}

{Thus, one can analytically determine the metric coefficients A(r) or B(r) via any specific assumptions. For instance, we can choose $A(r)=K_0/r$ and hence determine to $B(r)$ as follow.}

\begin{equation}\label{B(r)2}
 B= \frac{3}{4\alpha f_0 \sqrt{K_0}}(I_2 r^{1/2}-I_3 r^{3/2}).
\end{equation}

\begin{figure}
\resizebox{0.40\textwidth}{!}{\includegraphics{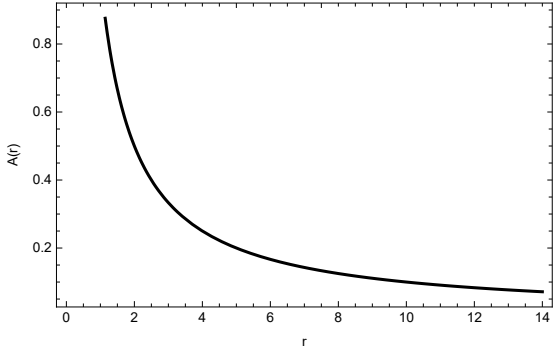}}
\caption{{Metric coefficient $A(r)$ as a function of the radial coordinate $r$ for the case of $w=-1/4$ with $\alpha=2$, $f_0=1/3$, $a_0=1$, $K_0=1$, $k_0=k_1=1$, $I_3=-1$ and $I_2=I_4=1$.}}
\label{fig:A(r)2}       
\end{figure}

\begin{figure}
\resizebox{0.40\textwidth}{!}{%
  \includegraphics{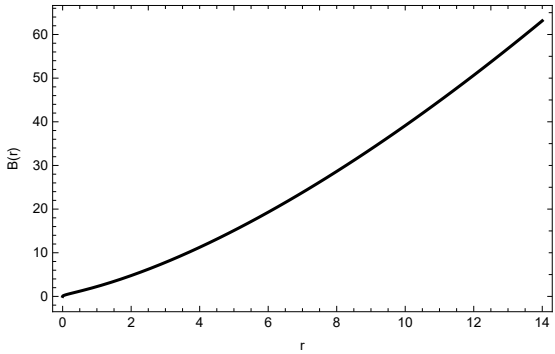}}
\caption{{Metric coefficient $B(r)$ as a function of the radial coordinate $r$ for the case of $w=-1/4$ with $\alpha=2$, $f_0=1/3$, $a_0=1$, $K_0=1$, $k_0=k_1=1$, $I_3=-1$ and $I_2=I_4=1$.}}
\label{fig:B(r)2}       
\end{figure}

{In Fig.\ref{fig:A(r)} and Fig.\ref{fig:B(r)}, for the case $w=-1/4$ the graphical behavior of the analytically determined expressions $A(r)$ and $B(r)$ are displayed, respectively. Similarly, the graphs of the metric coefficients determined outcome a specific selection on $A(r)$ for the case $w=1/5$ are exhibited in Fig.\ref{fig:A(r)2} and Fig. \ref{fig:B(r)2} As can be inferred, when the limit states are examined, when $r\rightarrow \infty$ $A(r)\rightarrow0$ and $B\rightarrow \infty$ for both cases. This behavior indicated that the asymptotical flatness.}

So far,  by taking $ f (R) = f_0R $ with a constant $ f_ 0 $, the situation General Relativity has been studied. In this case, the excess of Noether symmetries for the $ w = -1 / 4 $ and $ w = 1/5 $ values of the state equation parameter is striking, and this is clearly demonstrated above. These specific values of the $ w $ state equation parameter are different from known  values such as dust ($ w = 0 $), radiation ($ w = 1/3 $) and the cosmological constant ($ w = -1 $), etc. As an interpretation, it is possible to consider dark energy and dark matter as specific EoS values.

\section{Conclusions}

In this study, within the scope of $f(R)$ gravity theory, cosmological solutions for static cylindrical symmetric space-time were investigated and the Noether symmetry method was used to obtain exact solutions. As novel research compared to relevant studies in the past, we wanted to analyze cosmological solutions for a static cylindrical symmetric space-time, rather than for example studying compact objects using a spherically symmetric or BTZ background, in f(R) gravity.

We have constructed the theory and written the Lagrangian (\ref{lag}), where $\mathcal{L}_m\neq 0$ and the requirement that $w\neq 0$ is clearly seen from the Lagrangian. Then, the field equations of the theory have been obtained as (\ref{alan1}), (\ref{alan2}) and (\ref{alan3}).
The Noether symmetry method has been explained. Noether symmetry condition (\ref{NSkoş}) and first integral expressions (\ref{FirstInt}) have been given and their mathematical background has briefly been summarized.

Furthermore, the symmetry equations (\ref{ngs}) belonging to the theory have been determined. As a result of the constraints obtained from Noether symmetries, the specific values of $ w = -1 / 4 $ and $ w = 1/5 $ of the EoS parameter have explicitly been demonstrated. The Noether symmetries (\ref{s_ii}) and first integrals (\ref{int_ii}) have been found for the case of ${(i)}$ with $ w = -1 / 4 $. Here, the solutions (\ref{bulg_ii}) have directly been acquired by the first integrals. {The behavior of these solutions have been shown in Fig. \ref{fig:A(r)} and Fig. \ref{fig:B(r)}. For the case of ${(ii)}$ with $ w = 1/5 $, we have also studied Noether symmetries (\ref{s_iii}), first integrals (\ref{int_iii}), and hence the relation between the $A(r)$ and $B(r)$ (\ref{bulg_iii}) have been obtained. Then choosing the $A(r)=K_0/r$, we could obtained $ B(r)$ (\ref{B(r)2}). The behavior of this situations have also been depicted in Fig. \ref{fig:A(r)2} and Fig. \ref{fig:B(r)2}. Additionally, examining the limiting case of the solutions, one can get the asymptotic flatness.}

{This article is the first notice of an ongoing study that builds on $f(R)$ theory. The solutions are obtained in $f(R) = f_0 R$, that is, in GR theory. The general solutions obtained for the case $f(R)= f_0 R^n$ will be presented in the next article. Its difference from other solutions in the literature is that the values of the EoS parameter discussed in the solutions are not known. These values of the EoS parameter were obtained by considering the normally invisible constraints observed in the solution of the Noether symmetry equations. The solutions obtained in this way are new cosmological solutions and therefore this first part of the study is notable for being presented. We think that these new results are valuable because of the possibility that these determined values of the EoS parameter can express the structure of dark energy and dark matter.}
\section*{Acknowledgement}

The author Işıl Başaran Öz is supported by Istanbul University Post-Doctoral Research Project numbered MAB-2019-33386. The work of KB was supported in part by the JSPS KAKENHI Grant Number JP21K03547.

\begin{thebibliography}{99}

\bibitem{Perlmutter1998}
S.~Perlmutter \textit{et al.} [Supernova Cosmology Project],
Astrophys. J. \textbf{517}, (1999) 565-586.

\bibitem{Riess1998}
A.~G.~Riess \textit{et al.} [Supernova Search Team],
Astron. J. \textbf{116}, (1998) 1009-1038.

\bibitem{Weinberg1972}
S.~Weinberg,
\textit{"Gravitation and Cosmology: Principles and Applications of the General Theory of Relativity"}, (John Wiley and Sons, New York, 1972).

\bibitem{padmanabhan2007}
T.~Padmanabhan,
Gen. Rel. Grav. \textbf{40}, (2008) 529-564.

\bibitem{Copeland2006}
E.~J.~Copeland, M.~Sami and S.~Tsujikawa,
Int. J. Mod. Phys. D \textbf{15}, (2006) 1753-1936.

\bibitem{Durrer2008}
R.~Durrer and R.~Maartens,
Gen. Rel. Grav. \textbf{40}, (2008) 301-328.

\bibitem{Clifton2012}
T.~Clifton, P.~G.~Ferreira, A.~Padilla and C.~Skordis,
Phys. Rept. \textbf{513}, (2012) 1-189.

\bibitem{Bamba2012}
K.~Bamba, S.~Capozziello, S.~Nojiri and S.~D.~Odintsov,
Astrophys. Space Sci. \textbf{342}, (2012) 155-228.

\bibitem{Capozziello2011}
S.~Capozziello and M.~De Laurentis,
Phys. Rept. \textbf{509}, (2011) 167-321.

\bibitem{Nojiri2010}
S.~Nojiri and S.~D.~Odintsov,
Phys. Rept. \textbf{505}, (2011) 59-144.

\bibitem{Nojiri2017}
S.~Nojiri, S.~D.~Odintsov and V.~K.~Oikonomou,
Phys. Rept. \textbf{692}, (2017) 1-104.

\bibitem{Sotiriou2008}
T.~P.~Sotiriou and V.~Faraoni,
Rev. Mod. Phys. \textbf{82}, (2010) 451-497.

\bibitem{DeFelice2010}
A.~De Felice and S.~Tsujikawa,
Living Rev. Rel. \textbf{13}, (2010) 3.

\bibitem{Starobinsky1980}
A.~A.~Starobinsky,
Phys. Lett. B \textbf{91}, (1980) 99-102.

\bibitem{Carroll2003}
S.~M.~Carroll, V.~Duvvuri, M.~Trodden and M.~S.~Turner,
Phys. Rev. D \textbf{70}, (2004) 043528.

\bibitem{Nojiri2003}
S.~Nojiri and S.~D.~Odintsov,
Phys. Rev. D \textbf{68}, (2003) 123512.

\bibitem{Capozziello2003}
S.~Capozziello, S.~Carloni and A.~Troisi,
Recent Res. Dev. Astron. Astrophys. \textbf{1}, (2003) 625.

\bibitem{Bamba2009}
K.~Bamba and C.~Q.~Geng,
Phys. Lett. B \textbf{679}, (2009) 282-287.

\bibitem{Bamba2010a}
K.~Bamba and C.~Q.~Geng,
JCAP \textbf{06}, (2010) 014.

\bibitem{Bamba2011}
K.~Bamba, S.~Nojiri and S.~D.~Odintsov,
Phys. Lett. B \textbf{698}, (2011) 451-456.

\bibitem{Bamba2010b}
K.~Bamba, C.~Q.~Geng and C.~C.~Lee,
JCAP \textbf{11}, (2010) 001.

\bibitem{Bamba2013}
K.~Bamba, A.~N.~Makarenko, A.~N.~Myagky, S.~Nojiri and S.~D.~Odintsov,
JCAP \textbf{01}, (2014) 008.

\bibitem{Trendafilova2011}
C.~S.~Trendafilova and S.~A.~Fulling,
Eur. J. Phys. \textbf{32}, (2011) 1663-1677.

\bibitem{Sharif2012b}
M.~Sharif and S.~Arif,
Mod. Phys. Lett. A \textbf{27}, (2012) 1250138.

\bibitem{Houndjo2012}
M.~J.~S.~Houndjo, D.~Momeni and R.~Myrzakulov,
Int. J. Mod. Phys. D \textbf{21}, (2012) 1250093.

\bibitem{FarasatShamir2014}
M.~Farasat Shamir and Z.~Raza,
Astrophys. Space Sci. \textbf{356}, (2015) no.1, 111-118.

\bibitem{Sharif2014}
M.~Sharif and Z.~Yousaf,
Astrophys. Space Sci. \textbf{351}, (2014) 351-360.

\bibitem{Bhatti2017}
M.~Z.~u.~H.~Bhatti and Z.~Yousaf,
Annals Phys. \textbf{387}, (2017) 253-270.

\bibitem{Yousaf2020}
Z.~Yousaf, M.~Z.~Bhatti and T.~Naseer,
Phys. Dark Univ. \textbf{28}, (2020) 100535.

\bibitem{Momeni2017}
D.~Momeni, S.~Upadhyay, Y.~Myrzakulov and R.~Myrzakulov,
Astrophys. Space Sci. \textbf{362}, (2017) no.9, 148.

\bibitem{Azadi2008}
A.~Azadi, D.~Momeni and M.~Nouri-Zonoz,
Phys. Lett. B \textbf{670} (2008), 210-214.

\bibitem{Bronnikov2019}
K.~Bronnikov, N.~O.~Santos and A.~Wang,
Class. Quant. Grav. \textbf{37} (2020) no.11, 113002.

\bibitem{Capozziello2007}
S.~Capozziello, A.~Stabile and A.~Troisi,
Class. Quant. Grav. \textbf{24} (2007), 2153-2166.

\bibitem{Capozziello2012}
S.~Capozziello, N.~Frusciante and D.~Vernieri,
Gen. Rel. Grav. \textbf{44} (2012), 1881-1891.

\bibitem{Bahamonde2018}
S.~Bahamonde, K.~Bamba and U.~Camci,
JCAP \textbf{02} (2019), 016.

\bibitem{Camci2020}
U.~Camci,
Phys. Rev. D \textbf{103} (2021) no.2, 024001.

\bibitem{Darabi2013}
F.~Darabi, K.~Atazadeh and A.~Rezaei-Aghdam,
Eur. Phys. J. C \textbf{73} (2013), 2657.

\bibitem{Sawicki2007}
I.~Sawicki and W.~Hu,
Phys. Rev. D \textbf{75} (2007), 127502.

\bibitem{Sharif2012a}
M.~Sharif and S.~Arif,
Astrophys. Space Sci. \textbf{342} (2012), 237-243.

\bibitem{Momeni2009}
D.~Momeni and H.~Gholizade,
Int. J. Mod. Phys. D \textbf{18} (2009), 1719-1729.

\bibitem{Shamir2014}
M.~F.~Shamir and Z.~Raza,
Commun. Theor. Phys. \textbf{62} (2014) no.3, 348-352.

\bibitem{Capozziello1994}
S.~Capozziello and R.~De Ritis,
Nuovo Cim. B \textbf{109} (1994), 795-802.

\bibitem{Capozziello1996}
S.~Capozziello, R.~De Ritis, C.~Rubano and P.~Scudellaro,
Riv. Nuovo Cim. \textbf{19N4} (1996), 1-114

\bibitem{Capozziello2008}
S.~Capozziello and A.~De Felice,
JCAP \textbf{08} (2008), 016

\bibitem{Hussain2011}
I.~Hussain, M.~Jamil and F.~M.~Mahomed,
Astrophys. Space Sci. \textbf{337} (2012), 373-377

\bibitem{Jamil2011}
M.~Jamil, F.~M.~Mahomed and D.~Momeni,
Phys. Lett. B \textbf{702} (2011), 315-319

\bibitem{Kucuakca2011}
Y.~Kucuakca and U.~Camci,
Astrophys. Space Sci. \textbf{338} (2012), 211-216

\bibitem{Vakili2008}
B.~Vakili,
Phys. Lett. B \textbf{664} (2008), 16-20

\bibitem{Shamir2017}
M.~F.~Shamir and F.~Kanwal,
Eur. Phys. J. C \textbf{77} (2017) no.5, 286

\bibitem{Oz2017}
I.~B.~Oz, Y.~Kucukakca and N.Unal,
Can. J. Phys. \textbf{96} (2018) no.7, 677-680

\end{thebibliography}
\bibliographystyle{ieeetr}

\end{document}